\documentclass[lettersize,journal]{IEEEtran}
\IEEEoverridecommandlockouts
\usepackage{cite}
\usepackage{amsmath,amssymb,amsfonts}
\usepackage{graphicx}
\usepackage{textcomp}
\usepackage{xcolor}
\usepackage{float}
\usepackage{algpseudocode} 
\usepackage[ruled,linesnumbered]{algorithm2e}

\def\BibTeX{{\rm B\kern-.05em{\sc i\kern-.025em b}\kern-.08em
    T\kern-.1667em\lower.7ex\hbox{E}\kern-.125emX}}
\begin{document}

\title{Blockchain-Enabled Variational Information Bottleneck for Data Extraction Based on Mutual Information in Internet of Vehicles}
\author{Cui Zhang, Wenjun Zhang, Qiong Wu,~\IEEEmembership{Senior Member,~IEEE}, Pingyi Fan,~\IEEEmembership{Senior Member,~IEEE},\\
Nan Cheng,~\IEEEmembership{Senior Member,~IEEE}, Wen Chen,~\IEEEmembership{Senior Member,~IEEE}, and Khaled B. Letaief,~\IEEEmembership{Fellow,~IEEE}

\thanks{
Cui Zhang is with the School of Internet of Things Engineering, Wuxi Institute of Technology, Wuxi 214122, China (e-mail: faircas85@163.com).

Wenjun Zhang and Qiong Wu are with the School of Internet of Things Engineering, Jiangnan University, Wuxi 214122, China (e-mail: wenjunzhang@stu.jiangnan.edu.cn, qiongwu@jiangnan.edu.cn).

Pingyi Fan is with the Department of Electronic Engineering, Beijing National Research Center for Information Science and Technology, Tsinghua University, Beijing 100084, China (e-mail: fpy@tsinghua.edu.cn).

Nan Cheng is with the State Key Lab. of ISN and School of Telecommunications Engineering, Xidian University, Xi'an 710071, China (e-mail: dr.nan.cheng@ieee.org).

Wen Chen is with the Department of Electronic Engineering, Shanghai JiaoTong University, Shanghai 200240, China (e-mail: wenchen@sjtu.edu.cn).

K. B. Letaief is with the Department of Electrical and Computer Engineering, the Hong Kong University of Science and Technology (HKUST), HongKong (e-mail: eekhaled@ust.hk).
}}

\maketitle

\begin{abstract}
The Internet of Vehicles (IoV) network can address the issue of limited computing resources and data processing capabilities of individual vehicles, but it also brings the risk of privacy leakage to vehicle users. Applying blockchain technology can establish secure data links within the IoV, solving the problems of insufficient computing resources for each vehicle and the security of data transmission over the network. However, with the development of the IoV, the amount of data interaction between multiple vehicles and between vehicles and base stations, roadside units, etc., is continuously increasing. There is a need to further reduce the interaction volume, and intelligent data compression is key to solving this problem. The VIB technique facilitates the training of encoding and decoding models, substantially diminishing the volume of data that needs to be transmitted. This paper introduces an innovative approach that integrates blockchain with VIB, referred to as BVIB, designed to lighten computational workloads and reinforce the security of the network. We first construct a new network framework by separating the encoding and decoding networks to address the computational burden issue, and then propose a new algorithm to enhance the security of IoV networks. We also discuss the impact of the data extraction rate on system latency to determine the most suitable data extraction rate. An experimental framework combining Python and C++ has been established to substantiate the efficacy of our BVIB approach. Comprehensive simulation studies indicate that the BVIB consistently excels in comparison to alternative foundational methodologies.

\end{abstract}


\section{Introduction}
\subsection{Background}
With the development of vehicle intelligence and connectivity, the amount of data that vehicles need to process has increased sharply. The data includes sensor data, map information, user interaction data, etc., which puts forward higher requirements for the computing power of vehicles \cite{zhuang2019sdn, wang2024value}. However, limited by the hardware conditions and power consumption restrictions inside the vehicle, the computing power of a single car is insufficient, and the amount of data it can process is limited, usually failing to meet the computational services required by task demands \cite{luo2023edgecooper}. To solve this problem, it is necessary to rely on the Internet of Vehicles (IoV) networks. However, the rapid development of the IoV networks has attracted attackers to obtain users' private information by attacking vehicles, bringing challenges related to information security \cite{8976295}.

In the vehicle-to-vehicle network, the blockchain technology can establish a secure data link, addressing the issues of insufficient computing resources for each vehicle and the security of data transmission over the network \cite{cheng2019space}. When deploying blockchain technology, servers are typically set up to act as nodes in the blockchain network. These servers usually have sufficient computing resources and storage space, capable of processing large amounts of data and computational tasks \cite{zhang2022blockchain}. Servers collect various data from vehicles, package it into blocks. Each block contains a certain number of data records and is linked to another block, forming a reliable chain-like data structure and making it difficult to alter data within any block \cite{wu2022characterizing}. Therefore, the blockchain technology can ensure the security of the vehicle-to-vehicle network to a certain extent.

Nevertheless, as the vehicle fleet expands and the volume of data exchanges between vehicles and servers keeps rising, it is indeed necessary to further adopt effective data compression techniques to reduce the burden of data transmission and improve the overall efficiency and performance of the system \cite{sacone2024platoon}. The Information Bottleneck (IB) method, as a promising technology, play an important role in this field. Individually, vehicles initially assess the data's a priori likelihood, subsequently determining the data's mutual information grounded on this estimated probability, thereby gauging data interdependencies. Utilizing this, a codebook is engineered to optimize data mutual information, which is instrumental in crafting an encoding and decoding scheme \cite{tishby2000information}. The encoding process compresses the data, while the decoding retrieves its most valuable aspects, effectively downsizing the data and easing network traffic. However, the pursuit of maximizing mutual information introduces significant mathematical intricacies, making codebook development arduous \cite{10313285}. Moreover, the prerequisite probability estimation is contingent on inaccessible prior insights, further complicating mutual information computation.

Deep neural networks are capable of training encoder and decoder models without the prerequisite of codebook design, leading to the emergence of the Deep Information Bottleneck (DIB) approach \cite{wu2023characterizing}. Yet, the issue of calculating mutual information persists. To address this, the Variational Information Bottleneck (VIB) technique offers a solution by estimating mutual information through its upper and lower bounds, simplifying the issues encountered in DIB \cite{alemi2016deep}. Nonetheless, VIB's implementation of both encoder and decoder on a single device, compounded by the limited computational capabilities of vehicles, results in a significant computational load. Thus, the VIB algorithm has certain limitations for computationally intensive applications. Blockchain servers, equipped with ample computational power, offer an opportunity to offset the limited processing capabilities of vehicles by distributing the encoders and decoders between the vehicles and the servers. Consequently, there is a pressing need to develop a blockchain-enabled VIB procedure aimed at reducing the computational load and ensuring the security of vehicular networks.

\subsection{Related Work}
Certain studies have explored the application of blockchain solutions to alleviate the computational burden in the vehicle-to-vehicle network. Wei et al. designed a confusion strategy to offload encrypted computation to other devices, reducing the computational burden in the vehicle-to-vehicle network \cite{wei2021secure}. Nguyen et al. utilized smart contracts paired with double Q-networks to decrease the computational load on mobile devices within vehicle-to-vehicle networks, leveraging blockchain \cite{nguyen2021secure}. Lan et al. integrated drones into these networks, employing blockchain to ease computational demands \cite{10131971}. Yet, neither of these studies explored the application of VIB techniques for data compression within such networks.

In the realm of VIB applications, several methods have been put forward. Xie et al. developed a multimodal extension leveraging VIB for forecasting video content popularity \cite{9576573}. Shao et al. presented a novel communication framework using VIB to minimize transmission delays \cite{9606667}. Wang et al. merged self-supervised learning with VIB to tackle the challenge of limited samples in deep learning through pre-training on unlabeled data \cite{10158495}. Uddin et al. introduced a disentangled VIB approach combined with a Federated Learning loss function to curtail communication overhead \cite{9813696}. Li et al. showcased a VIB-based framework tailored for task-specific communications, enhancing accuracy and generalizability \cite{li2023task}. However, these endeavors have not tackled the computational and security issues inherent in VIB.

From our comprehensive review, no existing research has yet to integrate blockchain with VIB methods to simultaneously bolster network security and mitigate computational challenges, a gap that has propelled our current research endeavor.

\subsection{Contributions}
This paper introduces the blockchain-based Variational Information Bottleneck (BVIB) approach, aimed at reducing the computational load and safeguarding the security within vehicle-to-vehicle networks\footnote{The source code can be found at https://github.com/qiongwu86/BVIB-for-Data-Extraction-Based-on Mutual-Information-in-the-IoV}. Our key contributions include:

\begin{itemize}
	\item A novel network architecture that segments the model, with an encoder on devices and a decoder on servers, thereby alleviating the computational stress on vehicles.
	\item The introduction of an algorithm that integrates VIB with blockchain, marking the first such fusion and aiming to bolster network security without compromising VIB's decoding and compression capabilities.
	\item The development of an experimental platform integrating Python for VIB algorithm implementation and C++ for constructing a high-performance blockchain environment, interconnected through Dynamic Link Libraries (DLLs).
\end{itemize}

Considering the fluctuating environment of our autonomous vehicle network, characterized by changes in vehicle counts, server numbers, and wireless channel states, we have also formulated an online algorithm designed to curtail system latency under practical network scenarios.

\begin{itemize}
	\item On the basis of deriving the network delay formula, we use convex optimization methods to obtain the most suitable data extraction rate and design a dynamic optimization algorithm to minimize network latency.
\end{itemize}

Our work is the first attempt to combine blockchain with Variational Information Bottleneck. In the test results, both mutual information boundary accuracy and security have been optimized compared to other algorithms. The layout of this paper is organized as follows: Section II delineates the framework of the system model. Section III elaborates on the underlying algorithmic principles. Section IV offers an analysis of the experimental outcomes, while Section V concludes the paper.

\begin{figure*}[htbp]
	\vspace{-15pt}	
	\centerline{\includegraphics[width=0.8\textwidth]{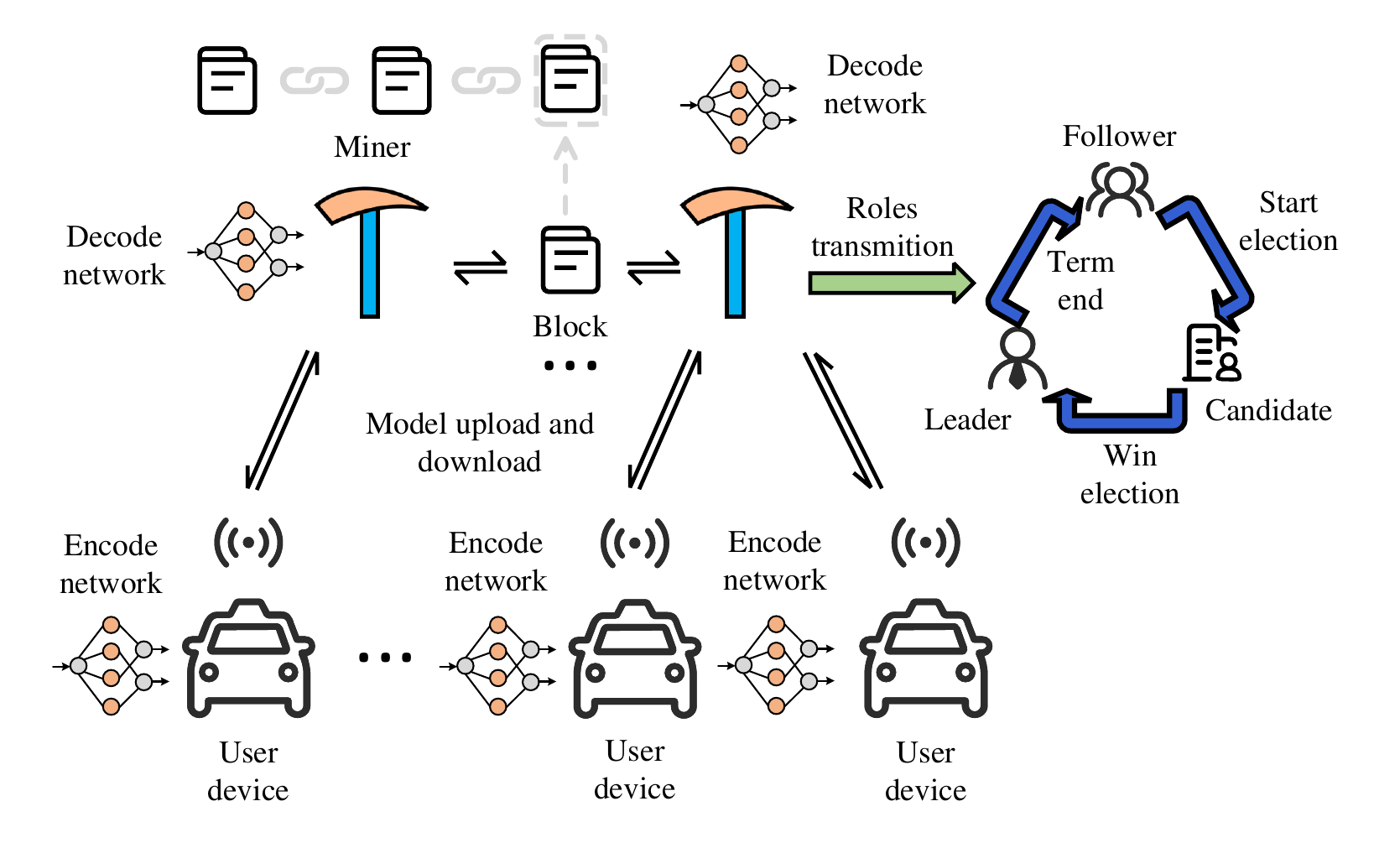}}
	\caption{Interaction between components of our blockchain-based varational bottleneck information approach for different oMVLs.}
	\label{fig1}
\end{figure*}
	
\section{System Model}
In this section, we construct a network architecture that integrates a multitude of vehicles with limited computing resources and servers with abundant computing resources (such as roadside units, base stations, etc.). Each vehicle communicates with a server. The server can change its identity between three roles (followers, candidates, and leaders), and all roles maintain a ledger of recorded information. The leader possesses the sole right to create blocks, while followers are tasked with documenting information and forwarding it to the leader. Candidates serve as intermediate roles transitioning from followers to leaders \cite{yuan2022trucon}. Each vehicle and server are equipped with an encoder and decoder in the form of neural networks, where the neural networks of the encoder and decoder are referred to as the encoding neural network and decoding neural network, respectively. Each vehicle is also loaded with a dataset that includes both training and testing data, segmented across several batches \cite{shen2023ringsfl}.

Data extraction by vehicles is treated as a Poisson process, leveraging its capability to represent a series of temporal events, thereby facilitating efficient forecasting and control of data traffic across the network \cite{sun2024knowledge}. The communication between vehicles and servers, which is suboptimal and aligns with the 3GPP LTE Cat. M1 standards, involves a set of servers, quantified by N, each capable of connecting to up to M vehicles. A visual representation of the system model can be found in Fig.~\ref{fig1}.

\section{BVIB process}
This section will provide a detailed explanation of the BVIB process, including mathematical derivations and the computation of its delay.
\subsection{Train stage}

\begin{algorithm}
	\SetKwData{Left}{left}\SetKwData{This}{this}\SetKwData{Up}{up}
	\SetKwFunction{Union}{Union}\SetKwFunction{FindCompress}{FindCompress}
	\SetKwInOut{Input}{input}\SetKwInOut{Output}{output}
	
	\Input{Require Data $X$ from dataset.}
	\BlankLine
	\emph{servers elect a leader}\;
	\For{$e\leftarrow 1$ \KwTo $E$}{
		\For{$i\leftarrow 1$ \KwTo $B$}{\label{forins}
			Vehicles input $X_i$ into the encoding network, obtaining $\mu$ and $\sigma^2$\; Vehicles send $\mu$ and $\sigma^2$ to nearby servers\;
			Servers use \eqref{reparameterization} to generate $\hat{Z_i}$\;
			Servers input $\hat{Z_i}$ into the decoding network to obtain $\hat{Y_i}$\;
			Servers calculate $\mathcal{L}_{\min}$ according to \eqref{eq123}\;
			Servers save $I(Z_i,X_i)_{max}$ and $I(Z_i,\hat{Y}_i)_{min}$ in the ledgers\;
			Adam algorithm is adopted to update the parameters of the encoding and decoding network\;
			The followers which are not attacked send their ledgers to the leader\;
			The leader collects the ledgers into a block and add it to the chain\;
		}
		Servers calculate $I(Z,X)_{max}$ and $I(Z,\hat{Y})_{min}$ according to \eqref{avm}\;
	}
	\caption{The training process of BVIB}
	\label{alg:bVIB}
\end{algorithm}

In the initial state, servers need to conduct an election to choose a leader. At the initial stage, all servers' roles are set as candidates, and each candidate randomly votes to elect a leader from among themselves. The candidate with the most votes is chosen as the leader, and the remaining candidates become followers, initially receiving random rewards. Upon the election of a leader, they initiate a genesis block to establish the chain and activate a timer for the routine dissemination of control messages throughout the duration $T_{term}$. Concurrently, servers prompt vehicles to commence training across $E$ epochs.

Subsequently, vehicles engage in data extraction, modeled as a Poisson process $\mathbb{R}$ for its efficacy in representing temporal event sequences, enabling the network to predict and regulate data traffic effectively. The arrival process A for vehicle data is characterized by:
\begin{itemize}
	\item The points of process $A([a,b))$ within any confined interval $[a,b))\in \mathbb{R}$ are described by a Poisson random variable with a mean rate $\lambda([a,b))$, treated as a non-negative Radon measure.
	\item For process A's points distributed across non-overlapping intervals $[a_1,b_1), [a_2,b_2),\cdots, [a_k,b_k)$, they constitute $k$ independent random variables with respective expected rates $\lambda([a_1,b_1)), \lambda([a_2,b_2)),\cdots, \lambda([a_k,b_k))$,  fulfilling the condition $a_k \leq b_k \leq a_{k+1}$ for $k=1,2,\cdots,K$.
\end{itemize}

For a point process over non-negative reals $\{{A_{(i)}}\}_{i\geq 1}$, points are sorted by their arrival times in ascending order $A_1\leq A_2\leq A_3\leq\cdots$. The inter-arrival time between consecutive points is a random variable $\mathcal{T}_i = A_i-A_{i-1}$, with $\mathcal{T}_1 = A_1$ for $i=2,3,4,\cdots$, introducing the concept of arrival gap or delay in our BVIB system. The two scenarios are as follows:

CASE \uppercase\expandafter{\romannumeral1} (Homogeneous Poisson Process): If $\{{A_{(i)}}\}_{i\geq 1}$ represents a homogeneous Poisson process with a constant data arrival rate $\Lambda$, the inter-arrival times  $\mathcal{T}_i$ are independent and identically distributed exponential variables with an average of $1/\Lambda$, given by:
\begin{equation}
	\mathbb{P}(\mathcal{T}_i \leq t) = 1 - e^{-\Lambda t},
	\label{eqa1}
\end{equation}

\begin{equation}
    \lambda(t) = \Lambda t.
    \label{eqa1.1}
\end{equation}

CASE \uppercase\expandafter{\romannumeral2} (Non-homogeneous Poisson Process): If $\{{A_{(i)}}\}_{i\geq 1}$ represents a non-homogeneous Poisson process with variable intensity $\Lambda(t)$, the arrival rate $\mathcal{T}_i$ is determined by the first arrival time $\mathcal{T}_1 = A_1$ with a distribution defined as:
\begin{equation}
	\mathbb{P}(\mathcal{T}_1 \leq t_1) = 1 - e^{-\int_0^{t_1}e^{-\Lambda (x)dx}}, \label{eqa2}
\end{equation}
then, considering the initial arrival delay $\mathcal{T}_1 = A_1$, the second conditional arrival time $\mathcal{T}_2$ is,
\begin{equation}
	\mathbb{P}(\mathcal{T}_2 \leq t_2|\mathcal{T}_1 \leq t_1) = 1 - e^{-\int_{t_1}^{t_2}e^{-\Lambda (x)dx}}, \label{eqa3}
\end{equation}
similarly for $i=3,4,\cdots$. In the non-homogeneous case $\{{A_{(i)}}\}_{i\geq 1}$, each point within the interva $a\in[0,t)$ is independently distributed with the following distribution:
\begin{equation}
	\mathbb{P}(A_i \leq a) = \frac{\lambda(a)}{\lambda(t)},
	\label{eqa4}
\end{equation}

\begin{equation}
	\lambda(t) = \lambda([0,t)) = \int_0^t\Lambda(t)dt,
	\label{eqa5}
\end{equation}
the cumulative rate of occurrence, denoted as $\lambda(t)$, is calculated by integrating the occurrence rate function $\Lambda(t)$ over time $0$ to time $t$. Here, $\Lambda(t)$ represents instantaneous arrival rate of vehicle data at time $t$. 

At the commencement of each epoch $e$ across the set $\{1,2,...,E\}$, the encoding and decoding network parameters are refined based on the prior epoch's outcomes. Vehicles then process each batch $B$ in sequence, with the $i$-th batch involving the input of $X_i$ into the encoding neural network, yielding the encoded dataset $Z_i$. Similar to \cite{alemi2016deep}, for any element $z$ within $Z_i$, it is hypothesized to follow a Gaussian distribution characterized by:
\begin{equation}
	p(z)=\mathcal{N}(z|\mu_i, \sigma^2_i),
	\qquad
	{\forall} z \in Z_i,
	\label{eq0}
\end{equation}
where the notation $\mathcal{N}$ signifies a Gaussian distribution, with ${\mu}_{i}$ and ${\sigma}^2_{i}$ denoting the mean and variance parameters that define the encoding function's distribution.

In line with the encoding principle outlined in reference \cite{alemi2016deep}, vehicles are capable of determining the parameters ${\mu}_{i}$ and ${\sigma}^2_{i}$ that align with equation \eqref{eq0}, and then transmit these to their designated servers. Thereafter, each server utilizes a reparameterization technique to formulate a new set of variables $\hat{Z}$, based on the received mean and variance values. For any element $\hat{z}$ within $\hat{Z_i}$, the formulation is as follows:
\begin{equation}
	\hat{z} = {\mu}_{i} + \epsilon{\sigma}^2_{i}, \qquad \forall \hat{z} \in \hat{Z},
	\label{reparameterization}
\end{equation}
with $\epsilon$ following a standard normal distribution $\mathcal{N}(0,1)$. It is evident from equations \eqref{eq0} and \eqref{reparameterization} that both $z$ and $\hat{z}$ adhere to an identical distributional pattern.

Subsequently, servers feed the reparameterized set $\hat{Z}_i$ into the decoding network to generate $\hat{Y}_i$ They then proceed to calculate an approximation of the loss function by determining its lower bound. With the assumption that $z$ and $\hat{z}$ are identically distributed, servers apply the VIB constraint to establish the bounds of mutual information: $I(Z_i,\hat{Y}_i)_{min}$ for the mutual information between $Z_i$ and $\hat{Y}_i$, and $I(Z_i,X_i)_{max}$ for that between $Z_i$ and $X_i$  Using these bounds, they compute the minimum value of the loss function as follows:
\begin{equation}
	\left\{
	\begin{aligned}
		&I(Z_i,\hat{Y}_i)_{min} = \sum_{n=1}^\mathbb{N}\sum_{m=1}^\mathbb{M} p(z_{n,i}|x_{m,i})\log q(\hat{y}_{m,i}|z_{n,i}), \\
		&I(Z_i,X_i)_{max} = \sum_{n=1}^\mathbb{N} \sum_{m=1}^\mathbb{M}p(z_{n,i}|x_{m,i})\log\frac{p(z_{n,i}|x_{m,i})}{r(Z_i)}, \\		
		&\mathcal{L}_{min} =  I(Z_i,\hat{Y}_i)_{min}	- \beta I(Z_i,X_i)_{max}, \\
		&{\forall} z_{n,i} \in Z_i, \qquad \forall \hat{y}_{i} \in \hat{Y}_i,
	\end{aligned}
	\right.
	\label{eq123}
\end{equation}
In this context, $\mathbb{M}$ and $\mathbb{N}$ represent the counts of data in batches $(X_i,\hat{Y}_i)$ and $Z_i$, respectively. The terms $x_{m,i}$ and $\hat{y}_{m,i}$ correspond to specific data points from $X_i$ and $\hat{Y}_i$, while $z_{n,i}$ is a data point from $Z_i$. The term $p(z_{n,i}|x_{m,i})$ signifies the conditional probability of $z_{n,i}$ conditioned on $x_{m,i}$, $q(\hat{y}_{m,i}|z_{n,i})$ is the variational approximation of the true conditional probability, and $r(Z_i)$ is the variational approximation of the marginal for $Z_i$. The Lagrange multiplier $\beta$ is introduced to achieve the minimum loss. This approach, in accordance with the Information Bottleneck principle, aims to maximize $I(Z_i,\hat{Y}_i)_{min}$ to match the decoded data $\hat{Y}_i$ with the target labels $Y_i$, and minimize $I(Z_i,X_i)_{max}$ to enhance data compression \cite{7133169}.

Following that, both vehicles and servers utilize the Adam optimization algorithm to adjust the weights of the neural networks for encoding and decoding. Specifically, the server dispatches gradient information from the initial layer to the vehicles, which then adjust the weights of the terminal layer accordingly \cite{kingma2014adam}. Upon gradient transmission, servers document the achieved $I(Z_i,X_i)_{max}$ and $I(Z_i,\hat{Y}_i)_{min}$ within their records.

Next, reliable followers transmit their records to the leader, while those compromised by attacks suffer a reduction in capability, preventing them from sending their records. Should the leader fail to gather at least half of the followers' records, the training sequence is aborted, and a fresh training phase is initiated. In contrast, if the leader successfully compiles all received records into a new block, this is appended to the blockchain, and the updated block is disseminated to all followers to synchronize their records. Thereafter, both the followers and the leader prompt the vehicles to feed the subsequent batch of data into the encoding network, thus iterating the training process until the total number of batches, denoted by $i$, equals $B$. At this juncture, the epoch's mean mutual information is determined by the equations presented below,
\begin{equation}
	\left\{
	\begin{aligned}
		&I(Z,X)_{max} = \frac{1}{B}\sum_{i=1}^{B}I(Z_i,X_i)_{max}, \\
		&I(Z,\hat{Y})_{min}=\frac{1}{B}\sum_{i=1}^{B}I(Z_B,\hat{Y_B})_{min}, \\		
	\end{aligned}
	\right.
	\label{avm}
\end{equation}
The iterative training continues until the minimum loss $\mathcal{L}_{min}$ stabilizes or the total epochs, represented by $e$, amount to $E$. The procedure is outlined in Algorithm \ref{alg:bVIB}.

It should be highlighted that if the leader encounters an attack or if the designated timer runs out during the training phase, the system will reset the timer, resume regular broadcasting of control messages, and trigger a new leadership election.

\vspace{-10pt}	
\subsection{Test Stage }
Compared with the stage of training, the testing phase does not include the parameter updating process.The system employs parameters that have been pre-trained and inputs the test dataset to obtain test results $\hat{Y}_i$. Servers record $\hat{Y}_i$ in the ledger instead of mutual information as in the training phase. The accuracy of each epoch is determined by the formula $(1-\frac{1}{B}\frac{1}{M}\sum_{i=1}^{B}\sum^M_{m=1}\hat{y}_{m,i}\oplus y_{m,i})\times100\%$, where $\oplus$ represents the bitwise exclusive OR operation between the predicted values $\hat{y}_{m,i}$ and the actual values $y_{m,i}$.
After the completion of the testing phase, time and CPU cycle consumption are recorded. Additionally, the average accuracy of testing is computed as the average across all epochs.

\vspace{-10pt}	
\subsection{Wireless Channel Model}
\begin{figure}[htbp]	
	\centerline{\includegraphics[width=0.5\textwidth]{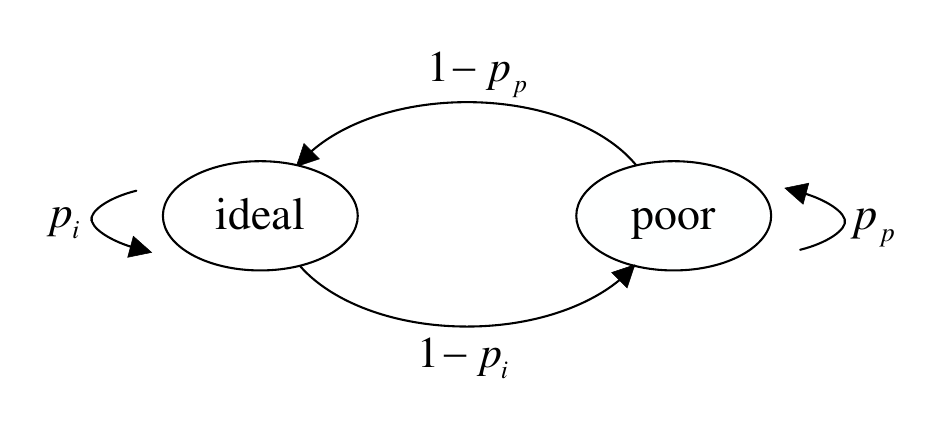}}
	\caption{Approximate Cellular Channel State Transition Diagram}
	\label{figcT}
\end{figure}

Communication delay is caused by data transmission between vehicles and servers. There are two distinct processes in the algorithm, namely uploading latent space vectors and backpropagation iterations. In our blockchain-driven framework, assessing communication latency, both for uploading and receiving feedback, necessitates a precise examination of the cellular network's channel dynamics, such as those for V2V and V2X LTE \cite{deng2023reconfigurable,yue2024hybrid}. We proceed with an approximated analysis that takes into account the channel state transitions as detailed in \cite{TMC_Pokhrel}, encapsulating the core concepts.

We operate under the premise of discrete time slots, with each slot's length equivalent to the transmission duration of a data block or module within our system. As depicted in Figure 3, the channel dynamics are presumed to alternate between optimal (where all transmissions are successful) and suboptimal states (resulting in transmission failures). In such a binary scenario, each transmission attempt has a fixed probability of success, aligning with a geometric distribution, thus offering only two potential outcomes: success or failure.

Define $\theta$ as the highest channel capacity for conveying link-layer frames to vehicles, calculated as the ratio of the bit rate to the frame size in bits. Let $v$ signify vehicle velocity, $f_c$ the carrier frequency, with the Doppler shift calculated by $f_d=f_cv/c$, where $c=3\times10^8m/s$. With F as the fade margin and a specific modulation and coding scheme in place, the channel is deemed poor if the received signal-to-noise ratio (SNR) is under the threshold $E[SNR]/F$, and ideal otherwise. The mean probability of frame transmission failure caused by channel errors is given by $\bar{p}_e=1-e^{-1/F}$. Additionally, $\eta=\sqrt{2/(F(1-\rho^2))}$, with $\rho=J_0(2\pi f_d/\theta)$, $\rho$ being the Gaussian correlation coefficient for the channel's fading amplitude at frequency $f_d$. The expression $1/\theta$ corresponds to the time taken to transmit a frame, and $J_0(.)$ refers to the zeroth-order Bessel function of the first kind. Ultimately, the channel's static transition probabilities are formulated utilizing a Markov model, i.e.,
\begin{equation}
	\left\{
	\begin{aligned}
		&P_p = 1 - \frac{\mathbb{Q}(\eta,\rho\eta)-\mathbb{Q}(\eta,\rho\eta)}{e^{1/F}-1}, \\
		&P_i = 1 - \frac{1-\bar{p}_e(2-P_p)}{1-\bar{p}_e}, \\		
	\end{aligned}
	\right.
	\label{pppi}
\end{equation}
The probabilities $P_p$ and $P_i$ respectively represent the probabilities of persisting in poor and ideal channels.

In \cite{TCOM_Pokhrel}, The probability of a single update being discarded in the channel is as follows.
\begin{equation}
	p_d=v_L\frac{1-p_i}{2-p_p-p_i}+\frac{l_L}{1+\frac{1-p_i}{1-p_p}},
	\label{pd}
\end{equation}
$v_L$ is the probability of one frame transmission failure in the poor channel, and $l_L$ is the probability of one frame transmission failure in the ideal channel.

\vspace{-10pt}	
\subsection{Delay computation}
First, vehicles need to extract data, which incurs an extraction delay $T_{ex}$. According to the property of homogeneous Poisson processes \eqref{eqa1}), the expectation of $T_{ex}$ is,
\begin{equation}
	E[T_{ex}]=\frac{1}{\lambda}.
	\label{tex}
\end{equation}

Taking into account the encoding delay $T_{ec}$, the delay of data when sent by vehicles is,
\begin{equation}
	T_m = T_{ex}+T_{ec}.
	\label{tm1}
\end{equation}

As one server is connected to M vehicles, there is a probability of collision among the data transmissions from these M vehicles. According to the LTE CAT M1 specification, collisions should be avoided by spacing transmissions 2 to 3 time slots apart. We adopt a spacing of 3 time slots, $\tau_c = 3\tau_t$, where $\tau_t$ is the time slot interval. Therefore, the probability of collision is given by:

\begin{equation}
	\begin{aligned}
		&p_c = 1-\prod^{M} Pr((T_{m_1}-T_{m_2})>\tau_c), \\
		&\forall m_1,m_2 \in [1,M] \quad\&\quad m_1 \neq m_2
		\label{pc1}
	\end{aligned}	
\end{equation}
Where $T_{m_1}$ and $T_{m_2}$ represent the times when any two vehicles send data. Also, due to \eqref{tm1}, $T_{m_1}$ and $T_{m_2}$ are Poisson process delays plus a constant, thus still constituting a Poisson process. Therefore, we have,
\begin{equation} 
	p_c = 1-e^{-\lambda M(M-1) \tau_c/2},
	\label{pc2}
\end{equation}
We assume that after a collision occurs, vehicles re-extract the data. Therefore, the updated delay for vehicle data transmission is,
\begin{equation}
	T_m = \frac{T_{ex}+T_{ec}}{1-p_c}.
	\label{tm2}
\end{equation}

Similarly, considering the probability of transmission failure in the channel (Eq. \ref{pd}), the delay for data to arrive at the server is,
\begin{equation}
	T_{ar} = \frac{T_m}{1-p_d} = \frac{T_{ex}+T_{ec}}{(1-p_c)(1-p_d)}.
	\label{tar}
\end{equation}

Then, the server decodes the data, resulting in a decoding delay $T_{dc}$, which is a constant. Followers upload decoded variables $\hat{Y}$ to the leader, giving a delay of $T_f$, also a constant. The leader generates and broadcasts the block to all followers, resulting in a delay of $T_p$, also a constant. Finally, the leader's election time $T_{ele}$ similarly affects the overall network delay.

Combining the above with \eqref{tar}, the overall network delay is given by,

\begin{equation}
	\begin{aligned}
		\mathbb{T} &= T_{ar}+T_{dc}+T_f+T_p \\
		    &= \frac{T_{ex}+T_{ec}}{(1-p_c)(1-p_d)} +T_{dc}+T_f+T_p+T_{ele}.
	\label{tn}
	\end{aligned}
\end{equation}

Combining \eqref{tex} and \eqref{pc2}, we have,

\begin{equation}
	E[\mathbb{T}] = \frac{1/\lambda+T_{ec}}{1-p_d}e^{\lambda M(M-1) \tau_c/2}+T_{si}+E[T_{ele}].
	\label{tt1}
\end{equation}
Where $T_{si} = T_{dc} + T_f + T_p$, which is for simplification purposes. $E[T_{ele}]$ represents the expected value of $T_{ele}$.

If there is no attack and each term is not interrupted, then the average election time overhead for one epoch is simply the expected value of $T_{ele}$, denoted as $E[T_{ele}]$.
\begin{equation}
	E[T_{ele}] = \frac{E[\mathbb{T}]}{T_{term}}\tau_{ele},
	\label{tele1}
\end{equation}
$\tau_{ele}$ is the time it takes to elect a leader. Assuming constant control information intervals, carrier frequencies, and other conditions, without loss of generality, we have $\tau_{ele}=\tau_b \log N$, where $N$ represents the server count and $\tau_b$ is a constant representing the election time.

For an attack with strength $a$ ($0 \leq a < \frac{N}{2}$), it means that out of $N$ servers, $a$ servers have been attacked and their computing power has been paralyzed. In our architecture, as long as the paralyzed servers are not the leader, the impact on the system is relatively small. If the leader is paralyzed, its control information stops, and an election will immediately begin. Assuming each server has an equal probability of being attacked, the probability of the leader being attacked is $\frac{a}{N}$. This results in additional election time overhead, so \eqref{tele1} is modified to be:
\begin{equation}
	E[T_{ele}] = \frac{E[\mathbb{T}]}{T_{term}}\tau_{ele}(1+\frac{a}{N}),
	\label{tele2}
\end{equation}

Combining \eqref{tele2}, \eqref{tt1} should be modified as follows,
\begin{equation}
	E[\mathbb{T}] = \frac{1/\lambda+T_{ec}}{1-p_d}e^{\lambda M(M-1) \tau_c/2}+T_{si}+\frac{E[\mathbb{T}]}{T_{term}}T_{ele}(1+\frac{a}{N}),
	\label{tt2}
\end{equation}

Combining equations \eqref{tele2} and \eqref{tt1}, we can revise as follows,
\begin{equation}
	E[\mathbb{T}] = \frac{NT_{term}}{NT_{term}-(N+a)T_{ele}}(\frac{1/\lambda+T_{ec}}{1-p_d}e^{\lambda M(M-1) \tau_c/2}+T_{si}),
	\label{tt3}
\end{equation}

\subsection{Dynamic optimization algorithm}

\begin{algorithm}
	\SetKwInOut{Input}{Input}\SetKwInOut{Output}{Output}
	\BlankLine
	\While{vehicles work}{
		\If{$\lambda \neq \lambda^*$}{
			$\lambda  \leftarrow \lambda ^*$ using \eqref{lamdapeak2}\;
		}
		
		\If{$N \rightarrow N^{new}$}{
			Jump step2\;
		}
		\If{$M \rightarrow M^{new}$}{
			Jump step2\;
		}
	}
	\caption{Real-time data extra algorithm}
	\label{alg:rda}
\end{algorithm}

The goal for the vehicles is to find the most suitable $\lambda$ to minimize $E[\mathbb{T}]$. Observing \eqref{tt3}, we find it to be a convex optimization function with respect to $\lambda$. We can differentiate it with respect to $\lambda$ to find the extreme points. Since \eqref{tt3} is quite complex, we can initially substitute $A=\frac{NT_{term}}{NT_{term}-(N+a)T_{ele}}$ and $B=M(M-1) \tau_c/2$ for these two coefficients.
\begin{equation}
	E[\mathbb{T}] = \frac{A}{1-p_d}[({T_{ec}}+\frac{1}{\lambda})e^{B\lambda}+T_{si}],
	\label{tt4}
\end{equation}
To take its derivative,
\begin{equation}
	E'[\mathbb{T}]=\frac{Ae^{B\lambda}}{(1-p_d)\lambda^2}(BT_{ec}\lambda^2+B\lambda-1),
	\label{ttd}
\end{equation}
Observing that $\frac{-AE^{-B\lambda}}{(1-p_d)\lambda^2}$ is always non-zero, $E'[\mathbb{T}]$ can only be zero when $BT_{ec}\lambda^2+B\lambda+1$ equals zero, for $E[\mathbb{T}]$ to attain an extreme value. According to Vieta's theorem,
\begin{equation}
	\lambda^*=\frac{-B\pm\sqrt{B^2+4BT_{ec}}}{2BT_{ec}},
	\label{lamdapeak1}
\end{equation}
Since $\lambda > 0$, \eqref{lamdapeak1} has only one valid solution,
\begin{equation}
	\lambda^*=\frac{-B+\sqrt{B^2+4BT_{ec}}}{2BT_{ec}},
	\label{lamdapeak2}
\end{equation}
Observing that within the interval $(0,\lambda^*)$, $E'[\mathbb{T}]$ is less than 0, hence $E[\mathbb{T}]$ decreases; and within the interval $(\lambda^*,+\infty)$, $E'[\mathbb{T}]$ is greater than 0, thus $E[\mathbb{T}]$ increases. Therefore, at $\lambda=\lambda^*$, $E[\mathbb{T}]$ reaches its minimum.

When a vehicle finds that its data extraction rate doesn't match $\lambda^*$, it adjusts its extraction rate to $\lambda^*$.

\section{PERFORMANCE EVALUATION}
In this section, we employ Python for our simulation environment, utilizing communication network configurations that adhere to the specifications of 3GPP LTE Cat. M1 rules\footnote{see technical reports at https://www.3gpp.org/DynaReport.}. We establish an experimental platform in this part and conduct simulation experiments by comparing with the baseline method to validate the efficiency of our proposed BVIB method. The experimental framework is established through the amalgamation of Python and C++, with Python tasked with executing the VIB algorithm and C++ utilized for developing a more efficient blockchain environment. Python and C++ programs are linked via Dynamic Link Libraries (DLLs). The adapted MNIST dataset from the National Institute of Standards and Technology, comprising 60,000 instances for training and 10,000 for testing, stands as a quintessential collection of data. We divide the dataset into batches, each with a size of 300. Detailed parameters are listed in Table \ref{tab1}.
\begin{table}[htbp]
	\caption{Parameter Settings.}
	\begin{center}
		\begin{tabular}{|c|c|c|}
			\hline
			\textbf{Description}&{\textbf{Value}} \\
			\hline
			$E$ & 300\\
			\hline
			$B$ & 200 \\
			\hline
			$M$ & 5\\
			\hline
			$N$ & 10\\
			\hline
			Learning rate & 0.001 \\
			\hline
			$T_{term}$ & 10min \\
			\hline
			$\tau_s$ & 1ms\\
			\hline
			$\tau_c$ & 3ms\\
			\hline
			$\tau_{ele}$ & 10ms\\
			\hline
			Maximum frequency of CPU & 5.20GHz\\
			\hline
			Number of vehicles & 15\\
			\hline
			Activate function & ReLU \\
			\hline
			Optimizer & Adam \\
			\hline
			Structure of encoder & 784-1024-512 \\
			\hline
			Structure of decoder & 512-784-10 \\

			\hline
		\end{tabular}
		\label{tab1}
	\end{center}
\end{table}
The baseline methods are outlined below:

(1) Variational Autoencoder (VAE): VAE is an unsupervised approach designed to distill concise representations from the input data. Given its esteemed status in the deep learning field, it has been selected as a benchmark for our comparative analysis.

(2) Conditional Variational Autoencoder (CVAE): CVAE expand upon VAE by incorporating conditional elements into the learning phase, enabling the creation of data aligned with particular criteria.

(3) VIB: VIB marks the pioneering effort in integrating variational inference with the information bottleneck concept, equipping it to manage intricate datasets and simplifying the computation of mutual information.

(4) Blockchain Federated Learning (BFL): BFL combines blockchain and federated learning, featuring distributed computation and decentralization, which can allow users to retain privacy data while achieving decent performance.

Consistent with the design of their respective encoder and decoder structures, vehicles and servers are allocated 2320 and 1306 neurons, respectively. For the subsequent experiments, the neuron count for VAE, CVAE, BFL, and VIB on each entity—whether vehicle or server—is standardized to match the aggregate neuron count of BVIB, totaling 3626. The outcomes presented are an average derived from a series of 10 trials.
\begin{table}[htbp]
	\caption{Computational resource comparison}
	\begin{center}
		\begin{tabular}{|c|c|c|c|}
			\hline
			\textbf{Device}&{\textbf{Neurons}}&{\textbf{Time cost}}&{\textbf{CPU cycles}} \\
			\hline
			VAE(vehicle) & 3626& 83.1s&254.73G \\
			\hline
			CVAE(vehicle) & 3626& 83.3s&261.12G \\
			\hline
			VIB(vehicle) & 3626& 83.2s&257.92G \\
			\hline
			BFL(vehicle) & 3626& 83.2s&258.32G \\
			\hline
			BFL(server) & 3626& 83.2s&258.32G \\
			\hline
			BVIB(vehicle) & 2320& 53.1s&164.61G  \\
			\hline
			BVIB(server) & 1306& 32.1s&99.51G  \\
			\hline
		\end{tabular}
		\label{tab2}
	\end{center}
\end{table}

Table \ref{tab2} illustrates the comparative resource usage of VAE, CVAE, VIB, BFL, and BVIB. It is evident that the proposed BVIB algorithm incurs less time and CPU cycle expenditure for both vehicles and servers compared to the other methods. This efficiency stems from the reduced neuron count assigned to vehicles in BVIB. The strategic distribution of the encoder on vehicles and the decoder on server nodes effectively diminishes the computational load on the vehicles.

\begin{figure}[htbp]
	\vspace{-10pt}		
	\centerline{\includegraphics[width=0.5\textwidth]{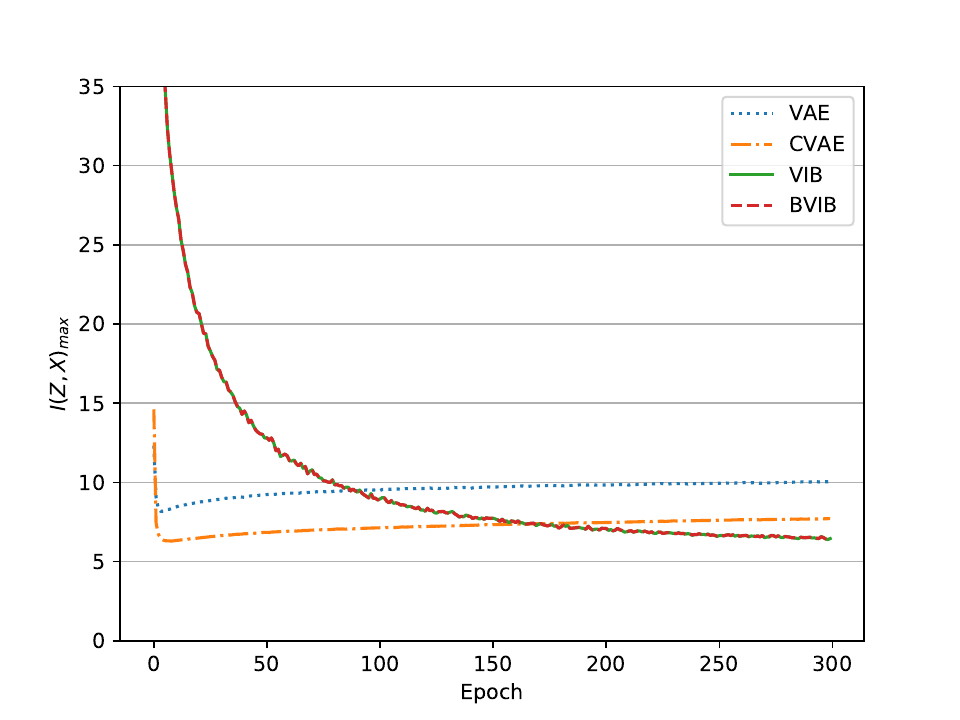}}
	\caption{$I(Z,X)_{max}$ under different approaches.}
	\label{fig2}
\end{figure}

Fig.~\ref{fig2} shows the progression of $I(Z,X)_{max}$ as iterations accumulate for four methodologies. VIB and BVIB display a consistent downward trajectory, surpassing VAE and CVAE in terms of reduction by the 240th iteration. This outcome is attributed to the fact that the loss functions for BVIB and VIB are specifically tailored to enhance mutual information optimization. On the other hand, VAE and CVAE solely account for cross-entropy during each iteration, neglecting mutual information. It's also observable that VIB and BVIB yield nearly the same $I(Z,X)_{max}$ values, signifying that BVIB's data compression performance closely matches that of VIB. Moreover, CVAE achieves a lower $I(Z,X)_{max}$ than VAE, due to its ability to use conditional data to improve compression efficiency.

\begin{figure}[htbp]
	\vspace{-15pt}	
	\centerline{\includegraphics[width=0.5\textwidth]{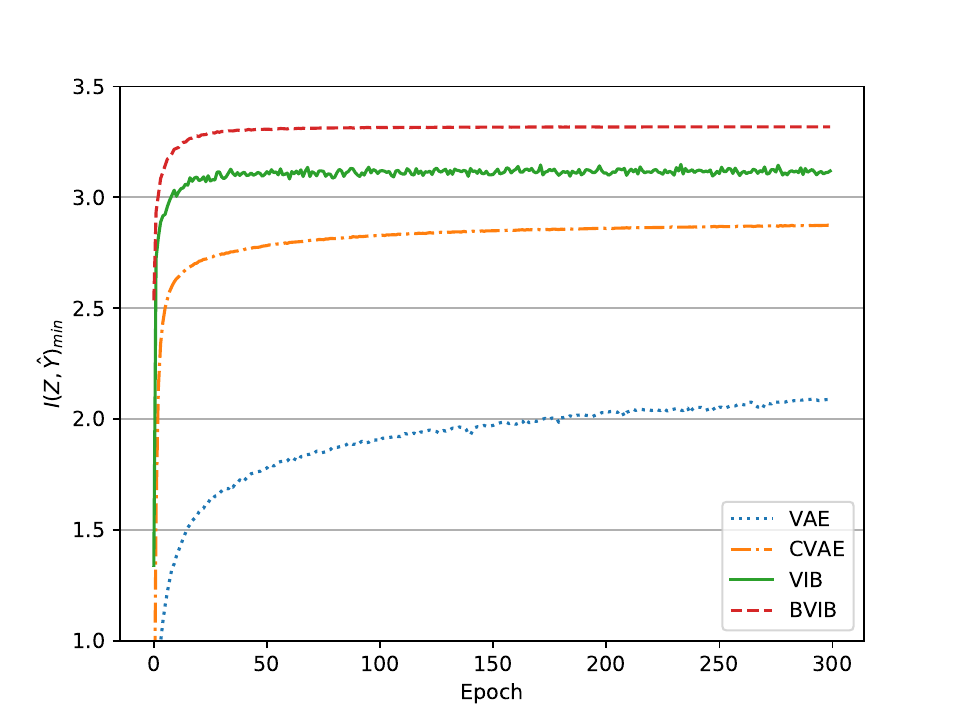}}
	\caption{$I(Z,\hat{Y})_{min}$ under different approaches.}
	\label{fig3}
\end{figure}
Fig.~\ref{fig3} demonstrates how $I(Z,\hat{Y})_{min}$ evolves as the iteration count increases using four distinct methods. It is evident that VIB and BVIB reach higher values than VAE and CVAE, due to their loss functions being engineered to maximize $I(Z,\hat{Y})_{min}$. Notably, the higher $I(Z,\hat{Y})_{min}$ under BVIB compared to VIB suggests BVIB's superior resilience to adversarial influences. Furthermore, it's observed that the convergence metric for VAE tends to be lower compared to CVAE. This can be attributed to the fact that VAE operates under an unsupervised paradigm, contrasting with the supervised approach of CVAE, where the latter typically exhibits superior training efficacy.

\begin{figure}[htbp]
	\vspace{-15pt}	
	\centerline{\includegraphics[width=0.5\textwidth]{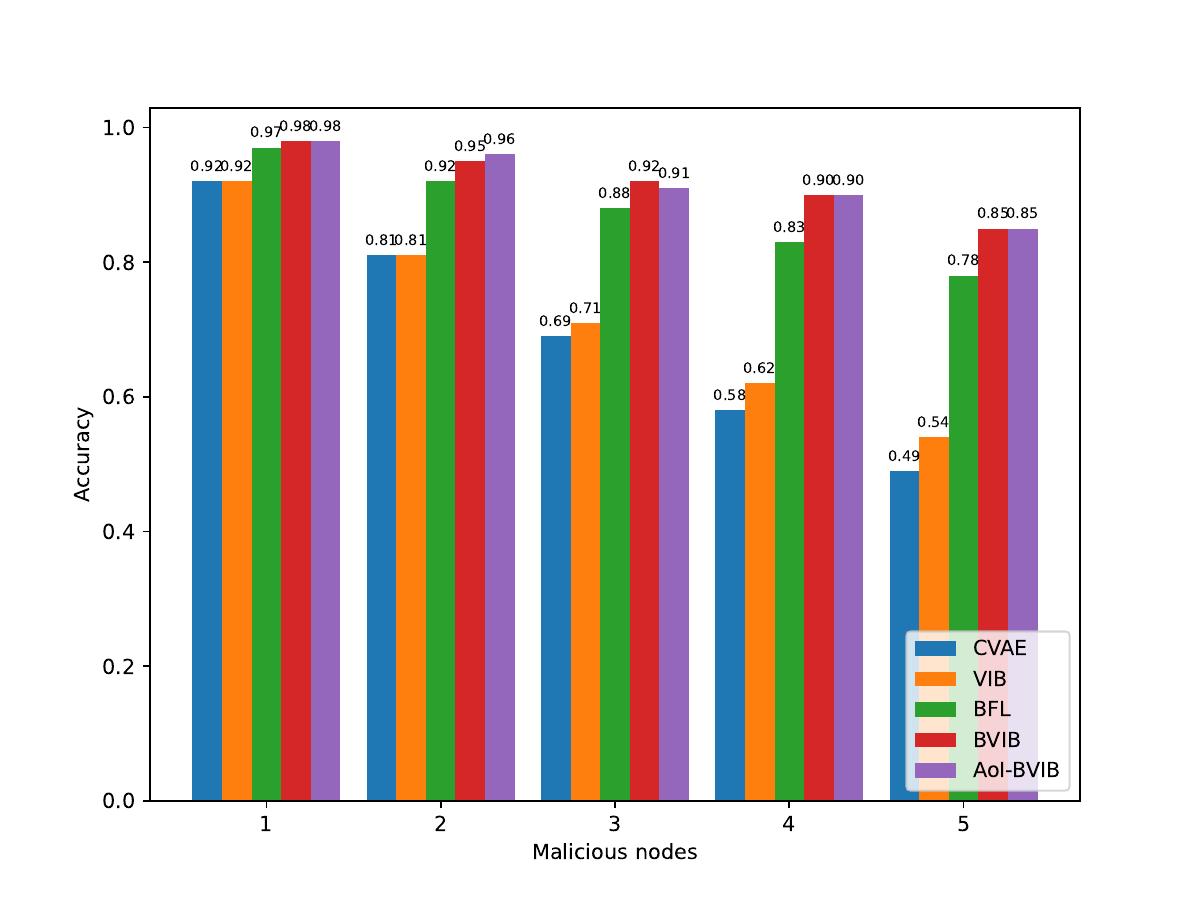}}
	\caption{Accuracy under different number of malicious nodes and approaches.}
	\label{fig4}
\end{figure}
Fig.~\ref{fig4} presents a comparison of the accuracy for VAE, CVAE, VIB, BFL, and BVIB in the presence of an escalating count of compromised nodes. It is observable that the accuracy for all methods declines with an increasing number of malicious nodes, as this results in a greater number of nodes being incapacitated due to attacks. Additionally, the accuracy of BFL and BVIB is much higher than that of VAE, CVAE, and VIB. This is because BFL and BVIB both incorporate blockchain to prevent malicious attacks. Moreover, BVIB has a slight edge over BFL due to its ability to swiftly elect a new leader upon an attack on the existing leader. Additionally, the Raft consensus algorithm employed by BVIB is deemed more efficient than the proof-of-work algorithm used by BFL.

\begin{figure}[htbp]
	\vspace{-10pt}
	\centerline{\includegraphics[width=0.5\textwidth]{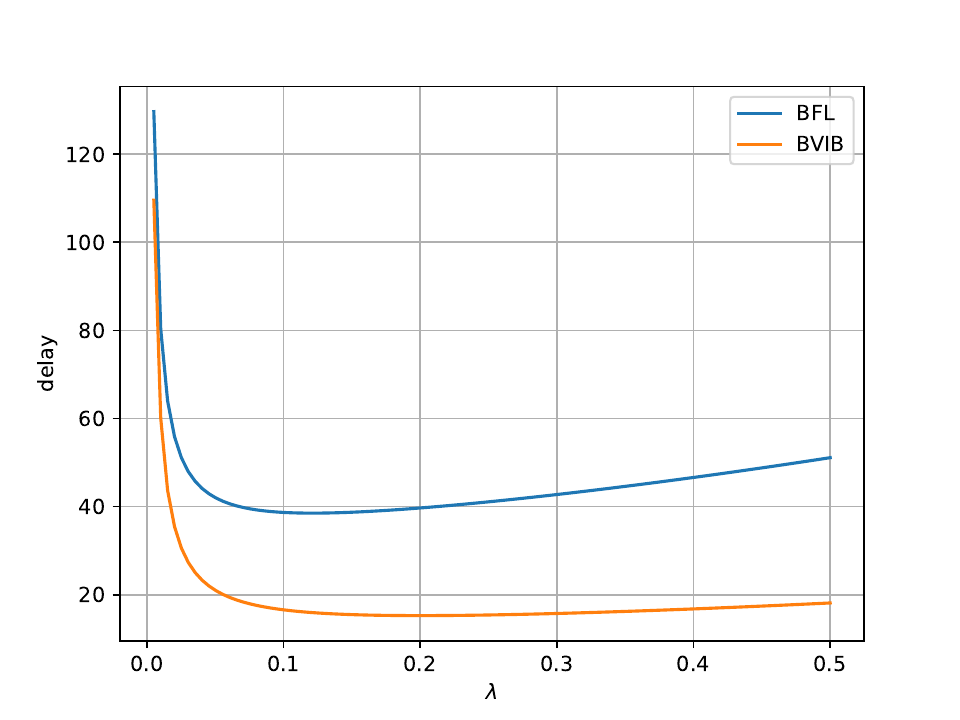}}
	\caption{Accuracy under attack.}
	\label{fig5}
\end{figure}
Fig.~\ref{fig5} compares the latency of BFL and BVIB with respect to varying Poisson arrival rates. In the figure, both exhibit a decreasing trend followed by an increase. However, overall, BVIB performs slightly better than BFL. This is because BFL employs the traditional consensus mechanism of proof-of-work, whereas in our work, we use the Raft mechanism, which brings significant optimization for blockchain latency. Additionally, because BFL transmits model parameters while our work involves transmitting latent space vector data, the scale of data transmission in BFL is considerably larger.

\begin{figure}[htbp]
	\vspace{-15pt}
	\centerline{\includegraphics[width=0.5\textwidth]{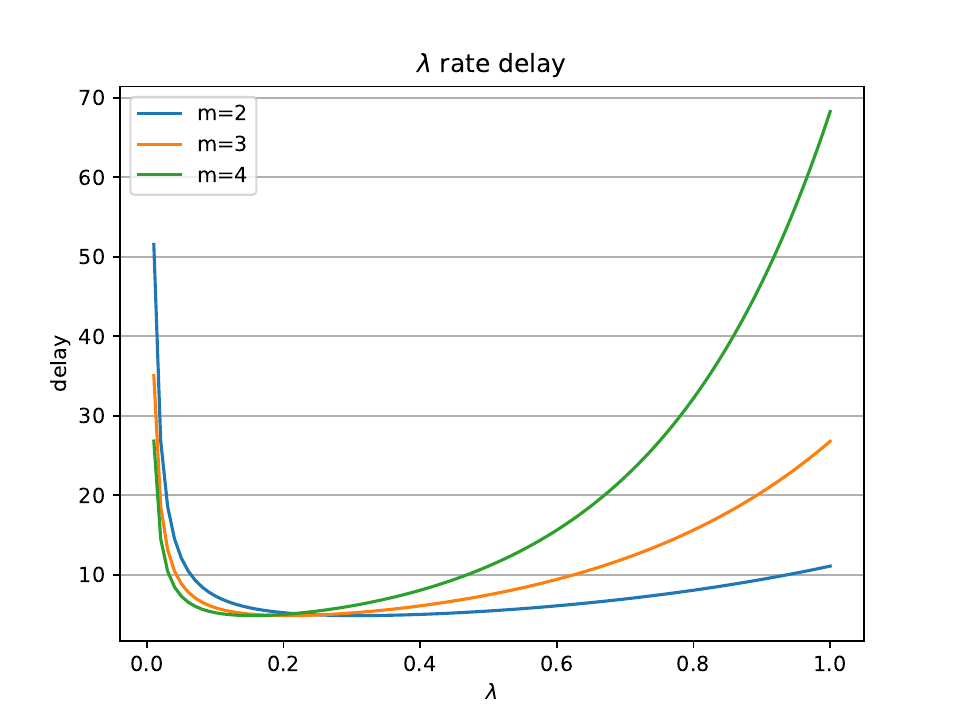}}
	\caption{latency with $\lambda$.}
	\label{fig6}
\end{figure}
Fig.~\ref{fig6} illustrates the relationship between system latency and the rate parameter $\lambda$. Depending on the different numbers of $m$, the optimal points of $\lambda$ vary. All three curves demonstrate that there exists an optimal point for system latency, which aligns with our theoretical derivations. In the case of $m=2$, the curve for $\lambda$ exhibits a rapid decrease followed by a rapid increase, with the optimal point at 0.22. Similarly, for $m=3$, the curve initially decreases rapidly then increases rapidly, with the optimal point at 0.18. In the case of $m=4$, the curve behaves similarly, with the optimal point at 0.16. This behavior arises from the properties of the formula for $\lambda$. Moreover, the minimum value of $\lambda$ when $m$ is 4 is smaller than the minimum value when $m$ is 2 or 3, indicating an appropriate value for the number of vehicles connecting to the same server.

\begin{figure}[htbp]
	\vspace{-10pt}	
	\centerline{\includegraphics[width=0.5\textwidth]{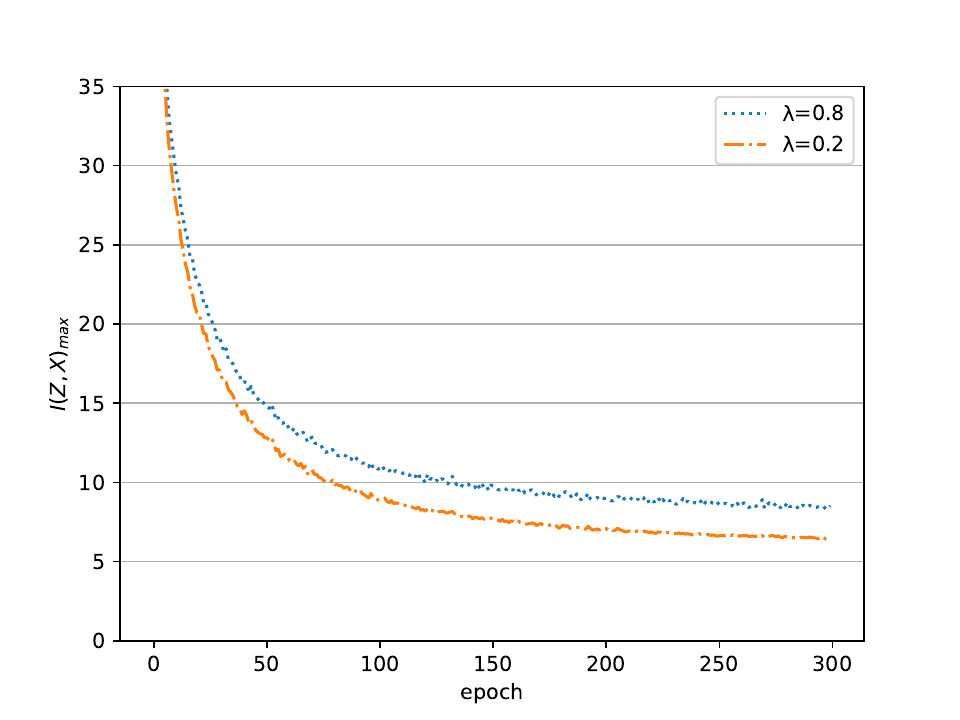}}
	\caption{$I(Z,X)_{max}$ with epoch.}
	\label{fig7}
\end{figure}
Fig.~\ref{fig7} depicts the relationship between the boundary value of $I(Z,X)_{max}$ and the number of iterations when $m=3$. With the progression of iterations, it's noticeable that the maximum value of $I(Z,X)_{max}$ diminishes and starts to stabilize around the 200th epoch. The patterns of the two curves remain largely the same, irrespective of whether the rate parameter $\lambda$ is configured at 0.2 or 0.8. Nonetheless, the system demonstrates improved performance when $\lambda$ is set at 0.2, correlating with $I(Z,X)_{max}$ reaching a lower maximum threshold. This is also determined by the properties of the formula for $\lambda$, where the optimal point for $\lambda$ when $m=3$ is around 0.2.
\begin{figure}[htbp]
	\vspace{-10pt}	
	\centerline{\includegraphics[width=0.5\textwidth]{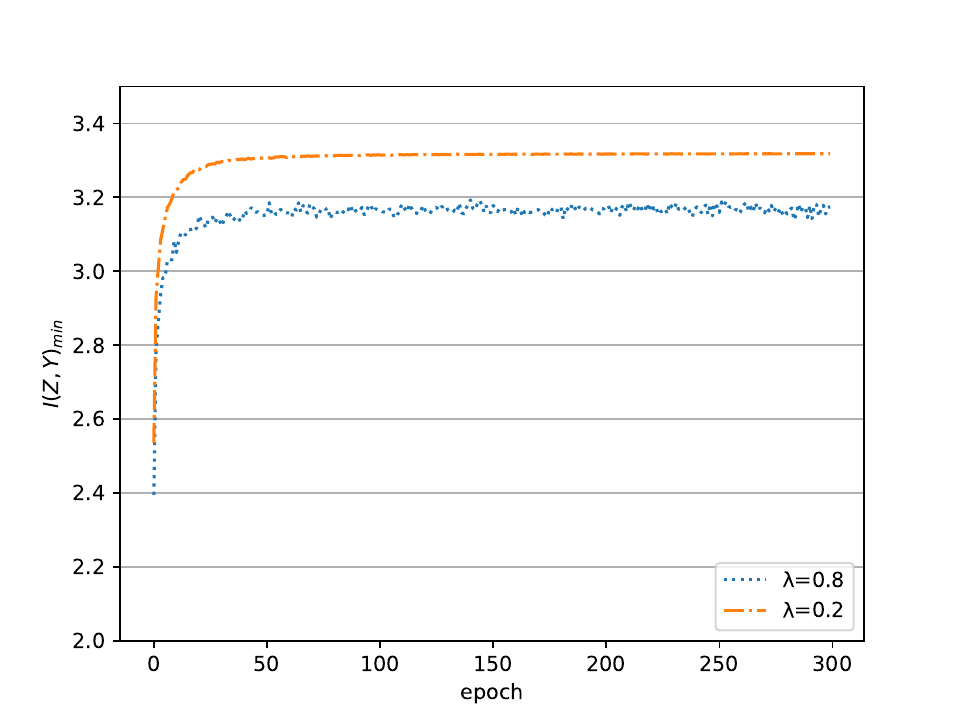}}
	\caption{$I(Z,\hat{Y})_{min}$  with epoch.}
	\label{fig8}
\end{figure}

Fig.~\ref{fig8} illustrates how the minimum boundary value of $I(Z,\hat{Y})_{min}$ correlates with the iteration count when $m$ is fixed at 3. It indicates an increase in iterations leads to an augmentation of the lower limit of $I(Z,\hat{Y})_{min}$. It is only around epoch 200 that the curve of $I(Z,\hat{Y})_{min}$ starts to flatten out. Regardless of whether $\lambda$ is set to 0.2 or 0.8, the trends of the two curves are almost the same. However, when $\lambda$ is 0.2, the overall system performs better, as $I(Z,\hat{Y})_{min}$ can achieve a larger upper bound. This is also determined by the nature of the $\lambda$ formula, where the optimal point for $\lambda$ when $m=3$ is around 0.2.

\begin{figure}[htbp]
	\vspace{-10pt}		
	\centerline{\includegraphics[width=0.5\textwidth]{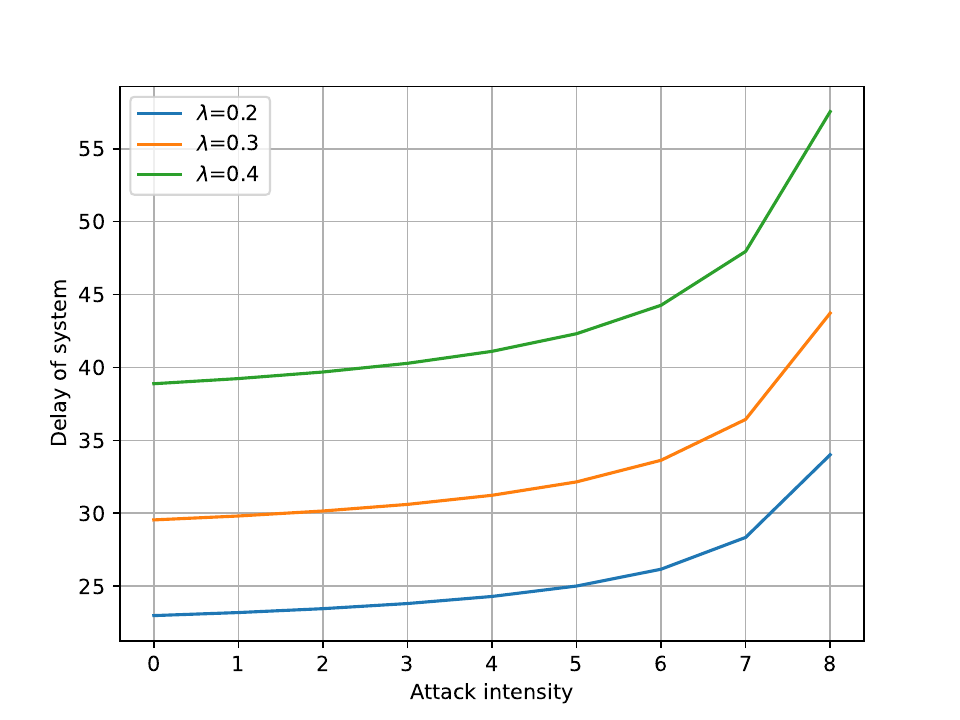}}
	\caption{Latency under attack.}
	\label{fig9}
\end{figure}
\begin{figure}[htbp]
	\vspace{-10pt}
	\centerline{\includegraphics[width=0.5\textwidth]{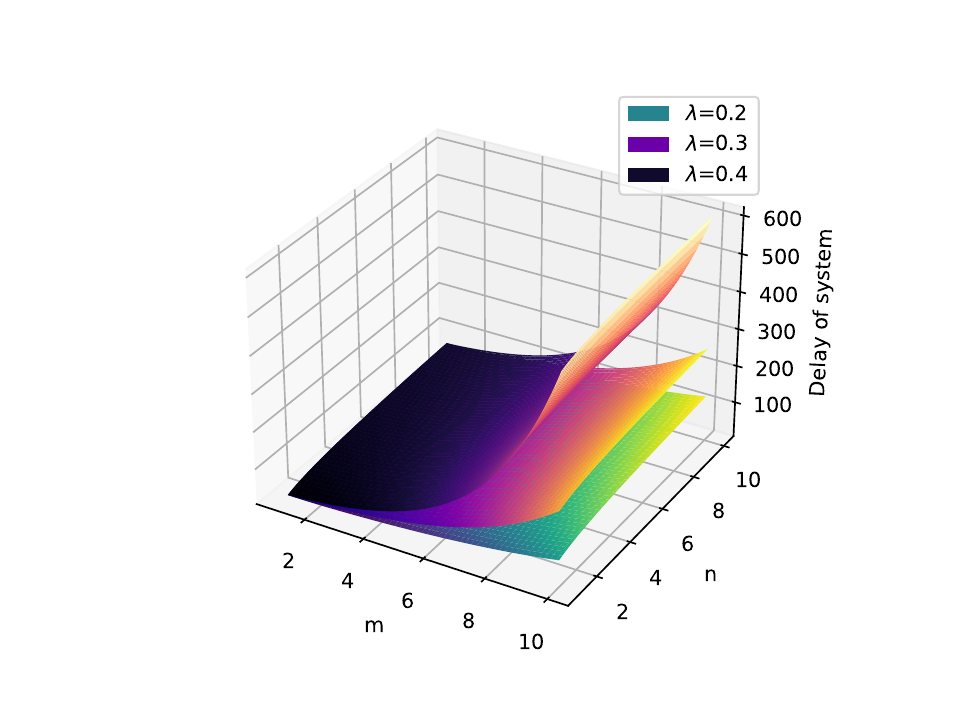}}
	\caption{Latency with $m$ and $n$.}
	\label{fig10}
\end{figure}
Fig.~\ref{fig9} illustrates the impact of attack intensity on system latency when $m=3$. The attack intensity is simulated by disabling servers, where an intensity of 1 represents disabling one server, and an intensity of 2 represents disabling two servers. The figure shows that as the attack intensity increases, the system latency also increases continuously. Before the attack intensity reaches 5, the increase in latency is not significant. However, after this point, the latency sharply rises, exhibiting a trend similar to an exponential function. A comparison between the two curves reveals that the overall performance is much better when $\lambda$ is set to 0.2 compared to when it is set to 0.3 or 0.4.

Fig.~\ref{fig10} illustrates the simultaneous impact of $m$ and $n$ on system latency. The graph demonstrates that the system's latency rises in conjunction with increases in both $m$ and $n$. However, the rate of increase in system latency accelerates with increasing $m$, while it gradually slows down with increasing $n$. This is due to the influence of the system latency formula. Similarly, the overall performance is much better when $\lambda$ is set to 0.2 compared to when it is set to 0.3 or 0.4.

\section{Conclusion}
In this article, we propose the BVIB method to alleviate the computational burden in vehicular networks and improve network security. We performed comprehensive simulation tests to showcase the capabilities of BVIB. The findings can be summarized as follows:
\begin {itemize}
\item Our proposed system architecture segregates the model, with the encoder operating on the device and the decoder on the server. This setup effectively alleviates the computational load on the vehicles.
\item The integration of VIB with blockchain in BVIB enhances the security of the vehicular network, concurrently maintaining the efficacy of the VIB.
\item We assessed how the rate of data extraction affects system latency and examined the implications of varying server counts and the maximum allowable vehicles connected. In real-time changing scenarios, we can use the optimization equation we calculate to adjust the data extraction rate and optimize the system delay.
\end {itemize}

\bibliographystyle{IEEEtran.bst}
\bibliography{ref}

\begin{thebibliography}{10}
\providecommand{\url}[1]{#1}
\csname url@samestyle\endcsname
\providecommand{\newblock}{\relax}
\providecommand{\bibinfo}[2]{#2}
\providecommand{\BIBentrySTDinterwordspacing}{\spaceskip=0pt\relax}
\providecommand{\BIBentryALTinterwordstretchfactor}{4}
\providecommand{\BIBentryALTinterwordspacing}{\spaceskip=\fontdimen2\font plus
\BIBentryALTinterwordstretchfactor\fontdimen3\font minus
  \fontdimen4\font\relax}
\providecommand{\BIBforeignlanguage}[2]{{%
\expandafter\ifx\csname l@#1\endcsname\relax
\typeout{** WARNING: IEEEtran.bst: No hyphenation pattern has been}%
\typeout{** loaded for the language `#1'. Using the pattern for}%
\typeout{** the default language instead.}%
\else
\language=\csname l@#1\endcsname
\fi
#2}}
\providecommand{\BIBdecl}{\relax}
\BIBdecl

\bibitem{zhuang2019sdn}
W.~Zhuang, Q.~Ye, F.~Lyu, N.~Cheng, and J.~Ren, ``{SDN/NFV}-empowered future
  {IoV} with enhanced communication, computing, and caching,''
  \emph{Proceedings of the IEEE}, vol. 108, no.~2, pp. 274--291, 2019.

\bibitem{wang2024value}
W.~Wang, N.~Cheng, M.~Li, T.~Yang, C.~Zhou, C.~Li, and F.~Chen, ``Value
  matters: A novel value of information-based resource scheduling method for
  {CAV}s,'' \emph{IEEE Transactions on Vehicular Technology}, 2024.

\bibitem{luo2023edgecooper}
G.~Luo, C.~Shao, N.~Cheng, H.~Zhou, H.~Zhang, Q.~Yuan, and J.~Li, ``Edgecooper:
  Network-aware cooperative lidar perception for enhanced vehicular
  awareness,'' \emph{IEEE Journal on Selected Areas in Communications}, 2023.

\bibitem{8976295}
J.~Cheng, G.~Yuan, M.~Zhou, S.~Gao, C.~Liu, H.~Duan, and Q.~Zeng,
  ``Accessibility analysis and modeling for {I}o{V} in an urban scene,''
  \emph{IEEE Transactions on Vehicular Technology}, vol.~69, no.~4, pp.
  4246--4256, 2020.

\bibitem{cheng2019space}
N.~Cheng, F.~Lyu, W.~Quan, C.~Zhou, H.~He, W.~Shi, and X.~Shen,
  ``Space/aerial-assisted computing offloading for {IoT} applications: A
  learning-based approach,'' \emph{IEEE Journal on Selected Areas in
  Communications}, vol.~37, no.~5, pp. 1117--1129, 2019.

\bibitem{zhang2022blockchain}
P.~Zhang, Y.~Wang, G.~S. Aujla, A.~Jindal, and Y.~D. Al-Otaibi, ``A
  blockchain-based authentication scheme and secure architecture for
  {IoT}-enabled maritime transportation systems,'' \emph{IEEE Transactions on
  Intelligent Transportation Systems}, vol.~24, no.~2, pp. 2322--2331, 2022.

\bibitem{wu2022characterizing}
F.~Wu, F.~Lyu, H.~Wu, J.~Ren, Y.~Zhang, and X.~Shen, ``Characterizing user
  association patterns for optimizing small-cell edge system performance,''
  \emph{IEEE Network}, vol.~37, no.~3, pp. 210--217, 2022.

\bibitem{sacone2024platoon}
S.~Sacone, ``Platoon control in traffic networks: New challenges and
  opportunities,'' \emph{IEEE Transactions on Intelligent Transportation
  Systems}, vol.~25, no.~4, pp. 149--183, 2024.

\bibitem{tishby2000information}
N.~Tishby, F.~C. Pereira, and W.~Bialek, ``The information bottleneck method,''
  \emph{arXiv preprint physics/0004057}, 2000.

\bibitem{10313285}
Z.~Li, R.~She, P.~Fan, C.~Peng, and K.~B. Letaief, ``Learning channel capacity
  with neural mutual information estimator based on message importance
  measure,'' \emph{IEEE Transactions on Communications}, 2023.

\bibitem{wu2023characterizing}
F.~Wu, F.~Lyu, J.~Ren, P.~Yang, K.~Qian, S.~Gao, and Y.~Zhang, ``Characterizing
  internet card user portraits for efficient churn prediction model design,''
  \emph{IEEE Transactions on Mobile Computing}, vol.~23, no.~2, pp. 1735--1752,
  2023.

\bibitem{alemi2016deep}
A.~A. Alemi, I.~Fischer, J.~V. Dillon, and K.~Murphy, ``Deep variational
  information bottleneck,'' \emph{arXiv preprint arXiv:1612.00410}, 2016.

\bibitem{wei2021secure}
X.~Wei, Y.~Yan, S.~Guo, X.~Qiu, and F.~Qi, ``Secure data sharing:
  Blockchain-enabled data access control framework for {I}o{T},'' \emph{IEEE
  Internet of Things Journal}, vol.~9, no.~11, pp. 8143--8153, 2021.

\bibitem{nguyen2021secure}
D.~C. Nguyen, P.~N. Pathirana, M.~Ding, and A.~Seneviratne, ``Secure
  computation offloading in blockchain based {I}o{T} networks with deep
  reinforcement learning,'' \emph{IEEE Transactions on Network Science and
  Engineering}, vol.~8, no.~4, pp. 3192--3208, 2021.

\bibitem{10131971}
X.~Lan, X.~Tang, R.~Zhang, W.~Lin, and Z.~Han, ``{UAV}-assisted computation
  offloading toward energy-efficient blockchain operations in internet of
  things,'' \emph{IEEE Wireless Communications Letters}, vol.~12, no.~8, pp.
  1469--1473, 2023.

\bibitem{9576573}
J.~Xie, Y.~Zhu, and Z.~Chen, ``Micro-video popularity prediction via multimodal
  variational information bottleneck,'' \emph{IEEE Transactions on Multimedia},
  vol.~25, pp. 24--37, 2023.

\bibitem{9606667}
J.~Shao, Y.~Mao, and J.~Zhang, ``Learning task-oriented communication for edge
  inference: An information bottleneck approach,'' \emph{IEEE Journal on
  Selected Areas in Communications}, vol.~40, no.~1, pp. 197--211, 2022.

\bibitem{10158495}
C.~Wang, S.~Du, W.~Sun, and D.~Fan, ``Self-supervised learning for
  high-resolution remote sensing images change detection with variational
  information bottleneck,'' \emph{IEEE Journal of Selected Topics in Applied
  Earth Observations and Remote Sensing}, vol.~16, pp. 5849--5866, 2023.

\bibitem{9813696}
M.~P. Uddin, Y.~Xiang, X.~Lu, J.~Yearwood, and L.~Gao, ``Federated learning via
  disentangled information bottleneck,'' \emph{IEEE Transactions on Services
  Computing}, vol.~16, no.~3, pp. 1874--1889, 2023.

\bibitem{li2023task}
H.~Li, C.~Zhu, Y.~Zhang, Y.~Sun, Z.~Shui, W.~Kuang, S.~Zheng, and L.~Yang,
  ``Task-specific fine-tuning via variational information bottleneck for
  weakly-supervised pathology whole slide image classification,'' in
  \emph{Proceedings of the IEEE/CVF Conference on Computer Vision and Pattern
  Recognition}, Vancouver, BC, Canada, 2023, pp. 7454--7463.

\bibitem{yuan2022trucon}
M.~Yuan, Y.~Xu, C.~Zhang, Y.~Tan, Y.~Wang, J.~Ren, and Y.~Zhang, ``{TRUCON}:
  blockchain-based trusted data sharing with congestion control in internet of
  vehicles,'' \emph{IEEE Transactions on Intelligent Transportation Systems},
  vol.~24, no.~3, pp. 3489--3500, 2022.

\bibitem{shen2023ringsfl}
J.~Shen, N.~Cheng, X.~Wang, F.~Lyu, W.~Xu, Z.~Liu, K.~Aldubaikhy, and X.~Shen,
  ``Ringsfl: An adaptive split federated learning towards taming client
  heterogeneity,'' \emph{IEEE Transactions on Mobile Computing}, 2023.

\bibitem{sun2024knowledge}
R.~Sun, N.~Cheng, C.~Li, F.~Chen, and W.~Chen, ``Knowledge-driven deep learning
  paradigms for wireless network optimization in 6{G},'' \emph{IEEE Network},
  2024.

\bibitem{7133169}
N.~Tishby and N.~Zaslavsky, ``Deep learning and the information bottleneck
  principle,'' in \emph{2015 IEEE Information Theory Workshop (ITW)},
  Jerusalem, Israel, 2015, pp. 1--5.

\bibitem{kingma2014adam}
D.~P. Kingma and J.~Ba, ``Adam: A method for stochastic optimization,''
  \emph{arXiv preprint arXiv:1412.6980}, 2014.

\bibitem{deng2023reconfigurable}
R.~Deng, Y.~Zhang, H.~Zhang, B.~Di, H.~Zhang, H.~V. Poor, and L.~Song,
  ``Reconfigurable holographic surfaces for ultra-massive {MIMO} in 6{G}:
  Practical design, optimization and implementation,'' \emph{IEEE Journal on
  Selected Areas in Communications}, vol.~41, no.~8, pp. 2367--2379, 2023.

\bibitem{yue2024hybrid}
S.~Yue, S.~Zeng, L.~Liu, Y.~C. Eldar, and B.~Di, ``Hybrid near-far field
  channel estimation for holographic {MIMO} communications,'' \emph{IEEE
  Transactions on Wireless Communications}, 2024.

\bibitem{TMC_Pokhrel}
S.~R. Pokhrel and M.~Mandjes, ``Improving multipath {TCP} performance over wifi
  and cellular networks: An analytical approach,'' \emph{IEEE Transactions on
  Mobile Computing}, vol.~18, no.~11, pp. 2562--2576, 2019.

\bibitem{TCOM_Pokhrel}
S.~R. Pokhrel and J.~Choi, ``Federated learning with blockchain for autonomous
  vehicles: Analysis and design challenges,'' \emph{IEEE Transactions on
  Communications}, vol.~68, no.~8, pp. 4734--4746, 2020.

\end{thebibliography}

\vfill

\vspace{12pt}

\end{document}